# Flexocoupling impact on the kinetics of polarization reversal

***Ivan S. Vorotiahin*[1, 2], *Anna N. Morozovska*[1], *Eugene A. Eliseev*[1, 3], *Yuri A. Genenko*[2*],**

[1] *Institute of Physics, National Academy of Sciences of Ukraine,*

*46, pr. Nauky, 03028 Kyiv, Ukraine*

[2] *Institut für Materialwissenschaft, Technische Universität Darmstadt,*

*Jovanka-Bontschits-Str. 2, 64287 Darmstadt, Germany*

[3] *Institute for Problems of Materials Science, National Academy of Sciences of Ukraine,*

*Krjijanovskogo 3, 03142 Kyiv, Ukraine*

## Abstract

The impact of flexoelectric coupling on polarization reversal kinetics and space charge dynamics in thin films of ferroelectric-semiconductors has been theoretically studied. The relaxation-type Landau-Khalatnikov equation together with the Poisson equation and the theory of elasticity equations have been used to calculate in a self-consistent way the spatial-temporal development of ferroelectric polarization, electric potential and space charge, elastic stresses, strains and their gradients. The analysis of the obtained results reveals a moderate increase of the flexocoupling influence on the polarization, elastic strain, electric potential and space charge distribution dynamics with the decrease of a ferroelectric film thickness. In contrast, the dependence of polarization switching time on the applied electric field is strongly affected by the flexocoupling strength. The polarization reversal process consists typically of two stages, the first stage has no characteristic time, while the second one exhibits a switching time strongly dependent on the applied electric field.

[*] genenko@mm.tu-darmstadt.de

The impact of flexoelectric effect ("flexoeffect"), which appears as elastic strain in response on the polarization gradient (and vice versa) [1, 2, 3], is of great importance for understanding and describing electrophysical and elastic properties of ferroics at nanoscale [4, 5, 6], such as ferroelectric thin films [7, 8, 9, 10], thin-film-based multilayer structures [11], nanoparticles [12, 13, 14], nanograin ceramics [15, 16] and nanocomposites [17]. This dependence is caused by a strong influence of electro-elastic field gradients on the nanoferroic properties [4-6], in contrast to macro-ferroics, where the gradients are pronounced only near surfaces and domain boundaries [18, 19, 20, 21, 22, 23].

This work is devoted to the theoretical study of the poorly understood influence of the flexoeffect on the polarization switching kinetics in thin ferroelectric films. In contrast to the moderate influence of flexoeffect on the thermodynamic polarization distribution, increasing monotonously with the film thickness decrease, its influence on the kinetics turns out to be unexpectedly strong, with a threshold at certain film thicknesses. These predicted phenomena have non-trivial physical nature and require in-depth experimental verification.

In a number of works the influence of the flexoeffect on phase transitions, thermodynamical equilibrium distributions of polarization, electric and elastic fields and the domain structures in nanoferroics was investigated [7-14, 18-22]. Theoretical conclusions are indirectly supported by experimental results [23, 24]. These results [7-14, 18-24] indicate the significant flexoeffect influence on all the thermodynamic characteristics of the ferroics. Flexocoupling values can be directly extracted from the experiment [25, 26, 27], they can be microscopically estimated [2] accounting for an upper limit from energetic considerations [28]. Otherwise the flexoelectric tensor elements can be calculated from the first principles [29, 30]. However, experimentally established parameters can differ from each other [25-27], as well as from the calculated values [2, 29-30] by several orders of magnitude [31]. The reasons for these discrepancies are still unclear. Dynamic flexoeffect [5, 32] has been investigated too poorly to speculate about potential impact on the polarization switching dynamics.

On the other hand, virtually all experimental investigations of electrophysical and electromechanical properties of ferroics were conducted in more or less non-stationary conditions. Domain wall pinning, caused by various defects of the crystal lattice, in most cases, makes ferroic to relax to the final equilibrium state during the time interval much greater than the experiment duration [33, 34, 35]. That is why investigating the impact of flexoeffect on the polarization reversal kinetics, distributions of elastic strains and stresses, electric fields and space charge is of the fundamental interest. Such theoretical research, as it appeared, is still lacking. That has motivated us to explore the flexocoupling impact on the kinetics of polarization reversal and space charge dynamics in thin films of the ferroelectrics with semiconducting properties.

The impact of the flexoelectric effect on the properties and structure of domain walls in ferroics is rather nontrivial, since the high gradients of the long-range order parameters (polarization vector components) and spontaneous strain components are always present in these regions [7-14, 18-22]. Hence, in thin ferroelectric films with domain structure there can be a complex interference of the polarization gradients caused by the several sources: domain walls, film surfaces and inhomogeneous strains via electrostriction and flexoelectric coupling ("flexocoupling"). Indeed, it becomes very difficult to establish the role of the flexocoupling due to the cross-talk of all aforementioned factors. Thus, in order to analyze the net flexocoupling impact on the polarization reversal kinetics in thin films, occurring from the system spatial confinement (i.e. from the presence of surfaces), here we lay aside the flexoelectric influence on the domain structures formation and growth. Therefore, all the results discussed below are valid for the single-domain polarization reversal in a thin film. The film thickness interval of 10 to 100 nm is considered. Of course, single domain switching in the 100 nm film is a theoretical abstraction; however, in the case of a 10 nm film, placed between the perfectly conducting electrodes, the stripe or closure domain structures can hardly occur for energy reasons in the case of low defect concentration [36] and in the absence of dislocations [37].

To find the spatial-temporal polarization distribution the relaxation Landau-Khalatnikov equation was used. It was solved numerically in a self-consisted manner together with the Poisson equation for electric potential and space charge distributions, and with the elasticity theory equations for the determination of elastic strains and stresses and their gradients.

Let us consider a film of ferroelectric-semiconductor with thickness $h$ sandwiched between parallel plane electrodes. We suppose that all physical quantities depend only on the distance $x_3$ from the upper electrode (1D problem). One-component ferroelectric polarization $P_3$ is normal to the film surface that corresponds to a tetragonal ferroelectric phase. Electric field $E_3$ is parallel to $P_3$ (see **Figure 1**).

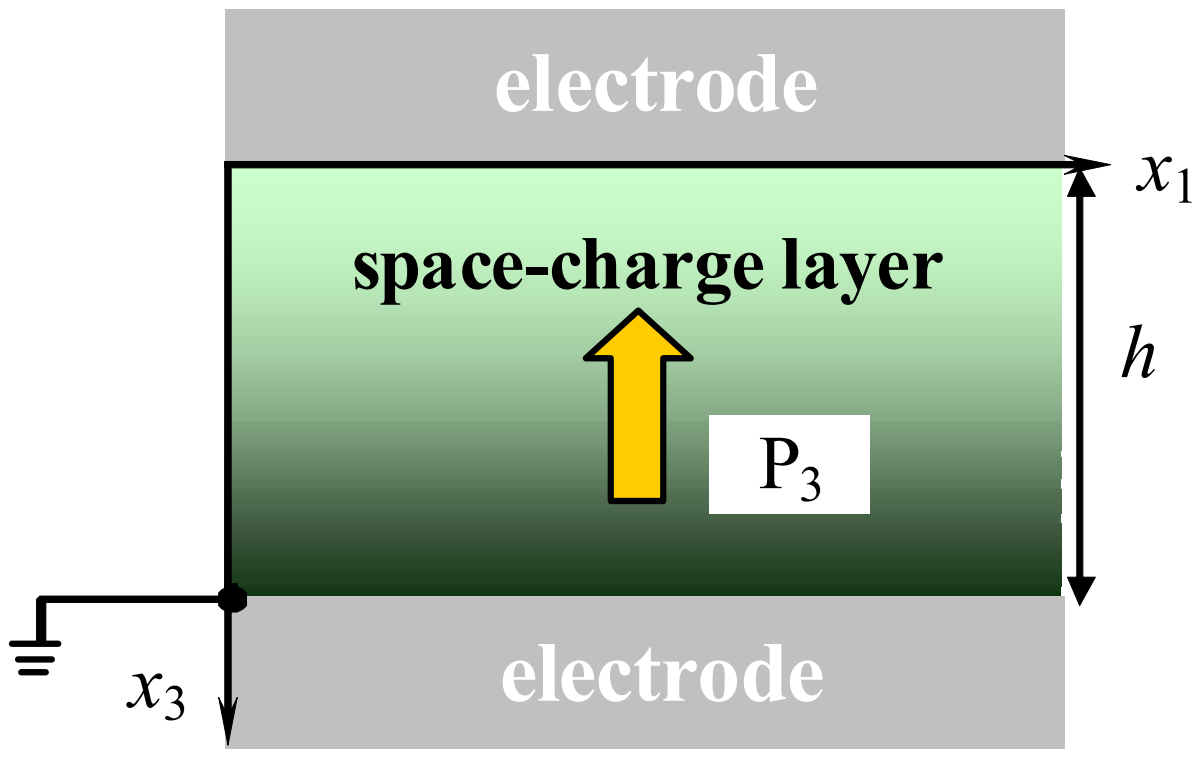

**Figure 1.** Schematics of the single-domain ferroelectric-semiconductor film with electrodes in the flat capacitor geometry. Polarization direction is shown by arrow.

In the considered 1D case, polarization vector has the form $\mathbf{P}=\left(0,0,P_3+\varepsilon_0\left(\varepsilon_b-1\right)E_3\right)$, where $\varepsilon_0$=8.85×10$^{-12}$ F/m is the dielectric permittivity of vacuum, $\varepsilon_b$ is the "background" dielectric permittivity not related to the ferroelectric polarization [38].

The charge carriers redistribution can create the internal electric field in the film, $E_3=-\partial\varphi/\partial x_3$, where the corresponding electric potential φ can be determined self-consistently from the Poisson equation resulting from the Gauss equation for the electrical displacement component $D_3=\varepsilon_0\varepsilon_b E_3+P_3$:

$$\varepsilon_0\varepsilon_b\frac{\partial^2\varphi}{\partial x_3^2}=\frac{\partial P_3}{\partial x_3}-e\left(Z_d N_d^+(\varphi)-n(\varphi)\right) \tag{1}$$

Here the electron density is $n$, ionized donor concentration is $N_d^+$, $e$=1.6×10$^{-19}$ C is the electron charge, $Z_d$ is the donor ionization degree (that is equal to zero for uncharged vacancies or isovalent impurities). The electric potential satisfies the boundary conditions at the electrodes, $\varphi(x_3=0,t)=U(t)$ and $\varphi(x_3=h,t)=0$. In the considered case the quasi-equilibrium donor concentration equals

$$N_d^+=N_d^0\left(1-f\left(\left(E_d+w_{ij}^d u_{ij}-eZ_d\varphi+E_F\right)/k_BT\right)\right), \tag{2a}$$

where the Fermi-Dirac distribution function is introduced as $f(x)=\left(1+\exp(x)\right)^{-1}$, $E_F$ is the Fermi energy level in equilibrium and $E_d$ is the donor level. Vegard expansion (another name is elastic dipole) tensor is $w_{ij}^d$ [39, 40]. The latter tensor will be regarded proportional to the unit tensor hereinafter, $w_{ij}^d=W\delta_{ij}$. The strain tensor components are $u_{ij}$. For the sake of simplicity we will use the quasi-equilibrium electron density as in the bulk ferroelectric,

$$n=N_C F_{1/2}\left(\left(e\varphi+E_F-E_C\right)/k_BT\right), \tag{2b}$$

where $N_C = \left(m_n k_B T/\left(2\pi\hbar^2\right)\right)^{3/2}$ is the effective density of states in the conduction band, electron effective mass is $m_n$ [41]. $F_{1/2}(\xi) = \frac{2}{\sqrt{\pi}}\int_0^\infty \frac{\sqrt{\zeta}d\zeta}{1+\exp(\zeta-\xi)}$ is the Fermi ½-integral that can be approximated using the formula (D.1) [42, 43] listed in the Suppl. Mat [44]. $E_C$ is the bottom of the conduction band.

Inhomogeneous spatial distribution of the ferroelectric polarization $P_3(x_3)$ can be determined from the time-dependent LGD (another name is Landau-Khalatnikov) equation:

$$\Gamma\frac{\partial}{\partial t}P_3 + a_{33}P_3 + b_{33}P_3^3 + \gamma_{333}P_3^5 - g_{33}\frac{\partial^2 P_3}{\partial x_3^2} + f_{kl33}\frac{\partial u_{kl}}{\partial x_3} - 2u_{kl}q_{kl33}P_3 = -\frac{\partial\varphi}{\partial x_3} \qquad (3)$$

where $\Gamma$ is the Khalatnikov coefficient, determined by the phonon relaxation time, $a_{33}(T) = \alpha_T\left(T - T_c\right)$, where $T$ is absolute temperature, $T_c$ is the Curie temperature of the bulk ferroelectric. The other coefficients of LGD thermodynamic potential are presented in Table C1 of **Appendix C** [44]. Corresponding boundary conditions read:

$$\left.\left(A_S P_3 - g_{33}\frac{\partial P_3}{\partial x_3} + f_{kl33}u_{kl}\right)\right|_{x_3=0} = 0 \text{ and } \left.\left(A_S P_3 + g_{33}\frac{\partial P_3}{\partial x_3} - f_{kl33}u_{kl}\right)\right|_{x_3=h} = 0. \qquad (4)$$

Here $A_S$ is the surface dielectric stiffness. The initial condition is $P_3(x_3, t=0) = 0$.

The equations of state, relating strain components, $u_{ij}$, to stress components, $\sigma_{ij}$, for a film, containing inhomogeneous distribution of ionized donors with concentration $\delta N_d^+ = N_d^+ - \overline{N}_d^+$ are

$$\sigma_{ij} = c_{ijkl}u_{kl} + w_{ij}\delta N_d + f_{ijkl}\frac{\partial P_k}{\partial x_l} - q_{ijkl}P_k P_l\ . \qquad (5)$$

Here $\overline{N}_d^+$ is the distribution of ionized donors in the absence of the applied voltage, $c_{ijkl}$, $f_{ijkl}$ and $q_{ijkl}$ are tensors of elastic stiffness, flexoelectric and electrostriction effects, respectively.

Equations (5) should be supplemented by the equilibrium conditions of bulk and surface forces, namely, $\partial\sigma_{ij}/\partial x_j=0$ in the bulk and $\sigma_{ij}n_j|_S=0$ at the free surface of the system. The static equation is valid under the realistic assumption that the propagation of polarization front is much slower the sound velocity. This leads to the equation for mechanical displacement vector $u_i$ inside the film:

$$c_{ijkl}\frac{\partial^2 u_k}{\partial x_i\partial x_j} - w_{kl}\frac{\partial\delta N_d}{\partial x_k} - f_{klmn}\frac{\partial^2 P_m}{\partial x_k x_n} - q_{klst}\frac{\partial\left(P_s P_t\right)}{\partial x_k} = 0. \qquad (6)$$

The boundary condition on the free surface of the film ($x_3 = 0$) is the absence of normal stresses, $\sigma_{3j}\left(x_3 = 0, t\right) = 0$. The surface $x_3=h$ is clamped to a rigid substrate and the displacement components are zero, $u_k\big|_{x_3=h}=0$. Expressions for elastic fields are presented in **Appendix A** [44]**.**

The coupled system of Eqs. (1), (3) and (6) with appropriate boundary conditions were rewritten in dimensionless variables and solved numerically (see **Appendix B** [44] for details). Obtained numerical results are presented and analyzed below.

The static and dynamic polarization, electric potential, donor and electron concentration distributions were calculated and analyzed for thin ferroelectric films of thickness (100 – 10) nm with and without taking into account the flexoeffect. Typical static distributions of the polarization and electric potential inside 100 nm and 10 nm films are shown in **Figure 2.** Corresponding donor and electron distributions are displayed in **Figures C1-C2** in the **Appendix C** [44]**.** With the decrease of film thickness we observe a moderate increase of the flexocoupling influence on the polarization distribution. The Vegard effect contribution turns out to be relatively small, so that one can neglect it.

The asymmetry of the polarization distribution is caused by the applied field, when its amplitude is high enough (10 MV/m and higher), and is also affected by asymmetric mechanical boundary conditions. The space charge distribution is also asymmetric since it follows the distribution of the electric potential mainly determined by the field applied to the film.

When the applied electric field is absent, this asymmetry is much smaller. The system becomes stable in the energetically favorable state with the spontaneous polarization value in the bulk region of the film. The amplitude of polarization drops rapidly from its maximum in the bulk towards zero near the film boundaries, driven by the boundary conditions. Electrostatic potential drops to zero at both film surfaces as demanded by boundary conditions (see **Appendix B** [44]). Two regions with the different sign of the potential are created within the bulk of the film. Flexocoupling makes these distributions asymmetric.

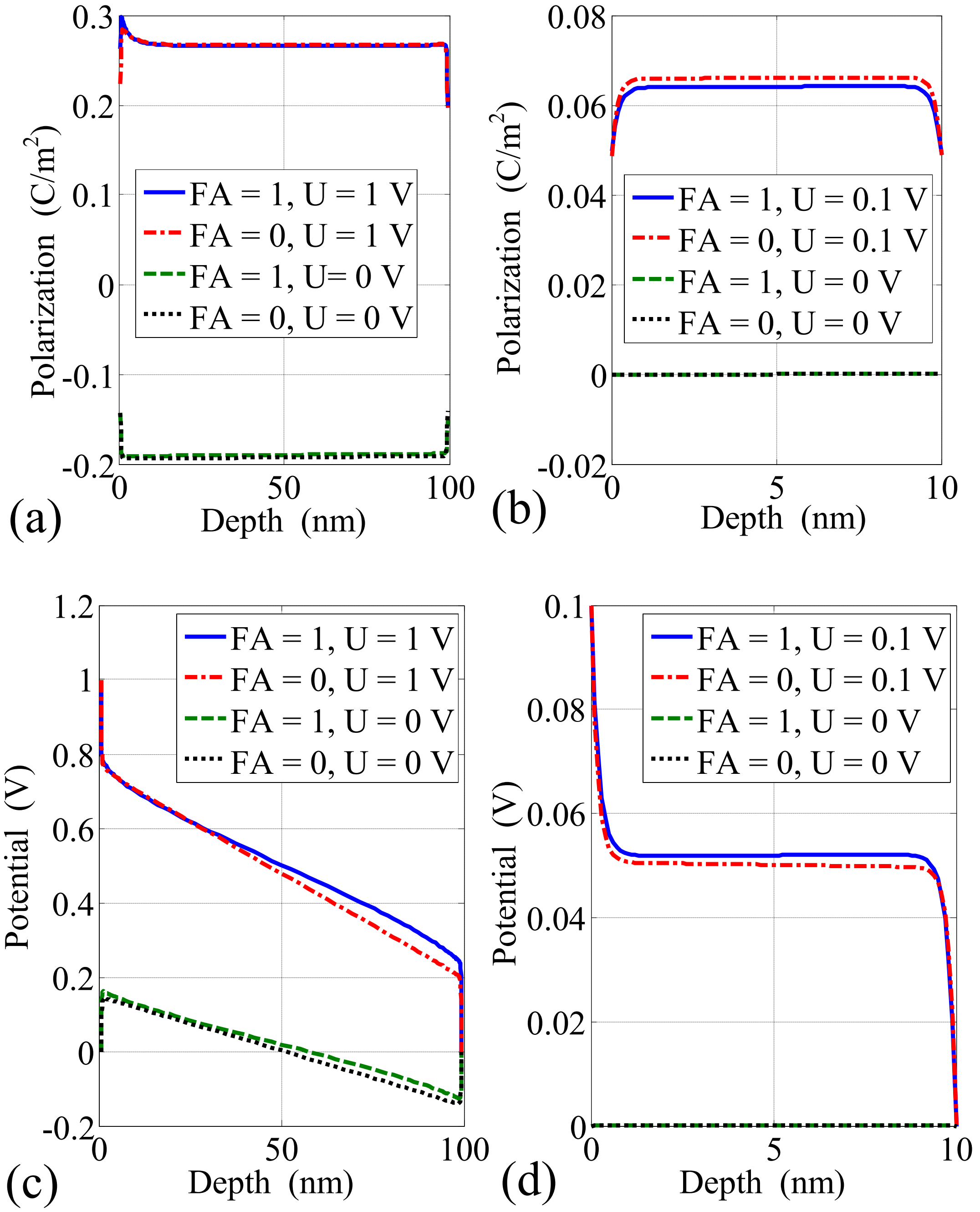

**Figure 2.** Spatial distributions of polarization **(a,b)** and electric potential **(c,d)** calculated for the films of 100 **(a,c)** and 10 nm **(b,d)** thickness. Different curves are calculated with (FA=1) and without (FA=0) flexocoupling, as well as with (U=1V, U=0.1V) and without (U=0) applied electric voltage. Material parameters are listed in **Table C1** [44]**.**

When considering kinetics of polarization switching the following scenario is studied. By a tiny voltage pulse the system is first allowed to polarize spontaneously in the negative direction. Then, after a while, the system is pulled out of the negative equilibrium state to the field-induced polarization in the positive direction by applying external electric field of the opposite sign. Corresponding polarization switching occurs generally in two steps: at first the system polarization insignificantly changes its value in the direction of the applied field and tries to become homogeneous across the film depth. As soon as it happens, the electric field rapidly switches all the system coherently to the state with the field-induced polarization. The first stage takes a while, but does not exhibit a characteristic time in the time dependence of the polarization (**Figure 3**). The second stage clearly reveals a characteristic switching time identified by the inflection point in the time dependence of the polarization (or by the peak in its time-derivative) and denoted by τ.

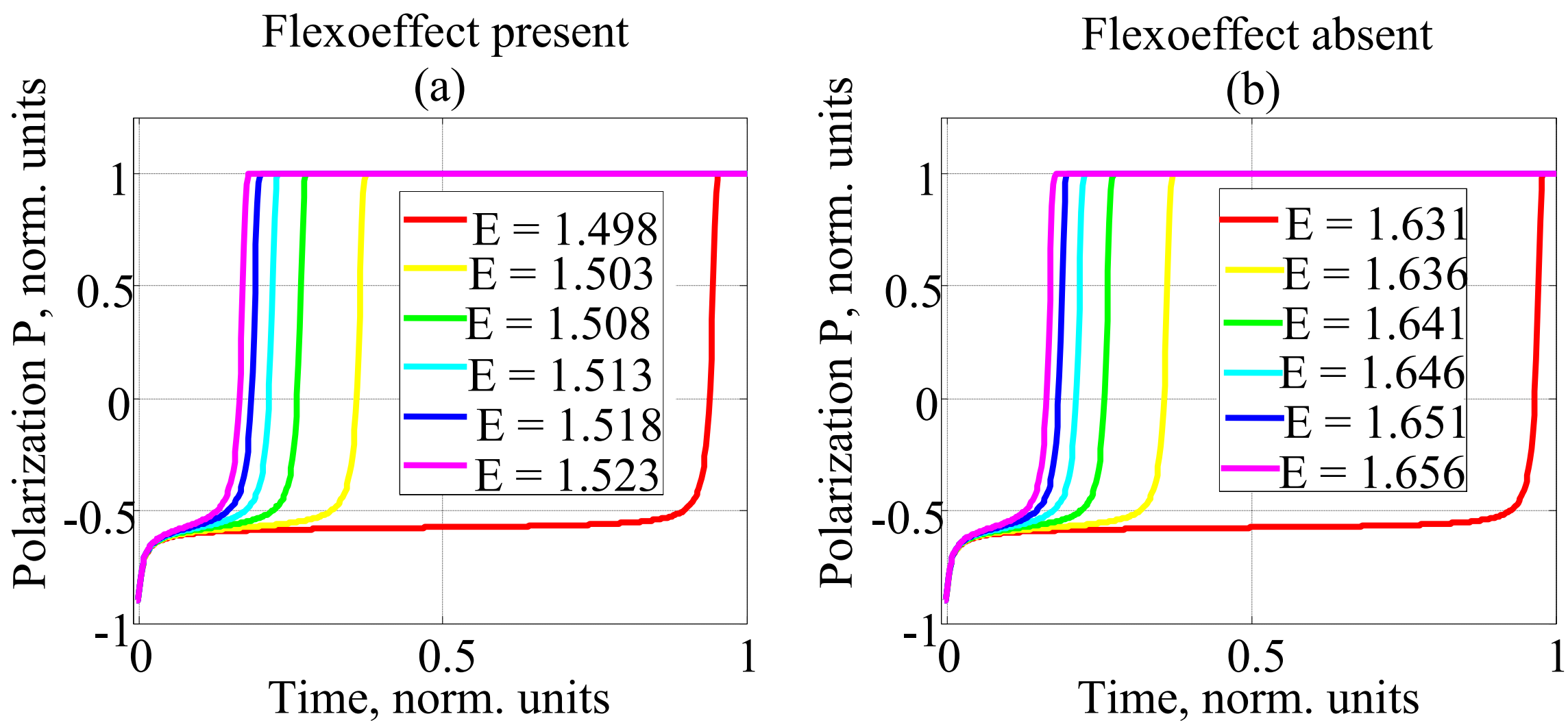

**Figure 3.** Single domain polarization reversal kinetics calculated for the 100 nm films with **(a)** and without **(b)** flexoelectric effect. Applied electric field is measured in MV/m. Parameters used in the modeling are listed in **Table C1**.

A homogeneous polarization distribution is much easier to achieve in the films of smaller thicknesses, so that the switching process occurs much faster there. A stable state of the system is close to the homogeneous distribution without applied field for the film thickness of $h_{cr}$ = 47 nm and smaller with flexocoupling (and for 46 nm without it), so that the system switches over the negligibly short time. In the films of bigger thickness the polarization reversal process has two stages, while the process occurs in one stage in the films of smaller thickness, as no further relaxation is required for the system to reach quasi-homogeneous "reversed" polarization distribution.

A well-known nontrivial difference of the single-domain switching from the multi-domain scenario is the existence of the critical field $E_{cr}$ below which the switching does not occur [45]. The critical field and the polarization reversal time drop down dramatically with the decrease of film thickness accompanied by a gradual reduction of spontaneous and field-induced polarization amplitudes. This trend can be explained by the analytical expressions for the thermodynamic coercive field for defect-free film (see eqs. (3) for in ref.[46] and eq. (12) in [47]): $E_c = \frac{2}{5}\left(2b_{33} + \sqrt{9b_{33}^2 - 20\alpha\gamma_{333}}\right)\left(\frac{2\alpha}{-3b_{33} - \sqrt{9b_{33}^2 - 20\alpha\gamma_{333}}}\right)^{3/2}$, where the coefficient $\alpha(T,h) \cong \alpha_T\left(T - T_c\left(1 - h/h_{cr}\right)\right)$, is renormalized by the finite size effects and screening conditions in thin films, and $h_{cr}$ is the critical thickness of the film size-induced transition into a paraelectric phase [48]. Note, that the lattice pinning phenomena, defects and elastic stresses can lead to the cross-over from the idealized "intrinsic" thermodynamic switching scenario to the realistic "extrinsic" switching one, based on domain nucleation by "favorable" defects at $E_{th} << E_c$, and hence to the actual thickness-independence of observable $E_c$ for the ultra-thin films [45].

Another nontrivial modeling result is the dependence of the characteristic time $\tau$ on the applied electric field which is strongly influenced by the flexoeffect (**Figure 4**). There is a narrow range in the electric field strengths, where the switching time $\tau$ goes down from the infinity to the values tending toward zero, the critical field $E_{cr}$ indicating the vertical asymptote of this dependence. For instance, in order to switch the 100 nm film during the one normalized time unit (which corresponds to the time span about $10^{-8}$ seconds, see **Appendix B [44]**) one needs to apply a critical field $E_{cr}$ of 1.5 MV/m to the film with account of the flexocoupling contribution or 1.6 MV/m without flexocoupling (see for details **Appendix C [44]**). The difference is about 10%, but the switching times of the process with and without flexocoupling contribution can be different by one order of the magnitude (see the vertical line in **Figure 4a**). Hence the flexoelectric effect facilitates the polarization switching and reduces the $E_{cr}$. The difference between the critical fields calculated with and without flexocoupling is $\Delta E_{cr} \approx 1.3\times10^5$ V/m. **Figure 4b** demonstrates that, in a wide range of the relative field strength variation, $(E\text{-}E_{cr})/\Delta E_{cr}$, the Landau approximation for the field dependence $\tau(E) \sim (E - E_{cr})^{-1/2}$ is valid.

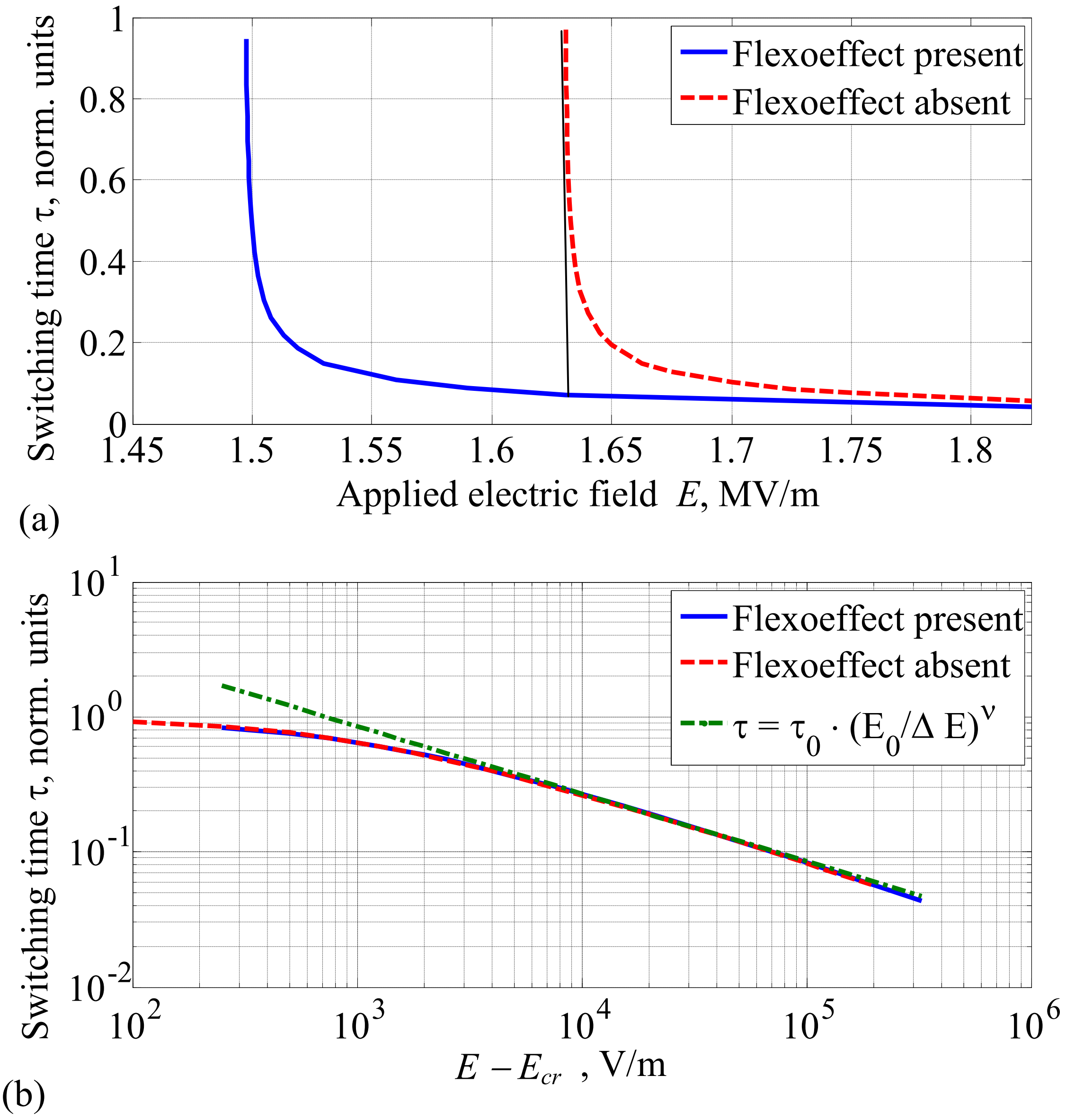

**Figure 4.** Dependence of the polarization switching time τ on the strength of applied electric field. Solid line denotes the dependences calculated with the flexocoupling ($E_{cr}^{fl}$ = 1.50 MV/m), dashed line shows the ones calculated without it ($E_{cr}^{nofl}$ = 1.63 MV/m), dotted line shows the asymptotic power behavior of both dependences. Film thickness is 100 nm. Material parameters are listed in **Table C1** [44].

Dependence of switching time on the relative amplitude of the flexoelectric coefficient is shown in **Figure 5a**. Thickness dependence of the critical field is presented in **Figure 5b**, where the thickness threshold ($h_{cr}$) is clearly seen. The physical nature of this phenomenon is related with the strong flexocoupling influence on the polarization dynamics and elastic strain gradient distribution near the film surfaces. After all an explicit dependence of the polarization kinetics on the strength of the flexoeffect is displayed in **Figure 6** for the 100nm film subject to a fixed external field of 1.631 MV/m.

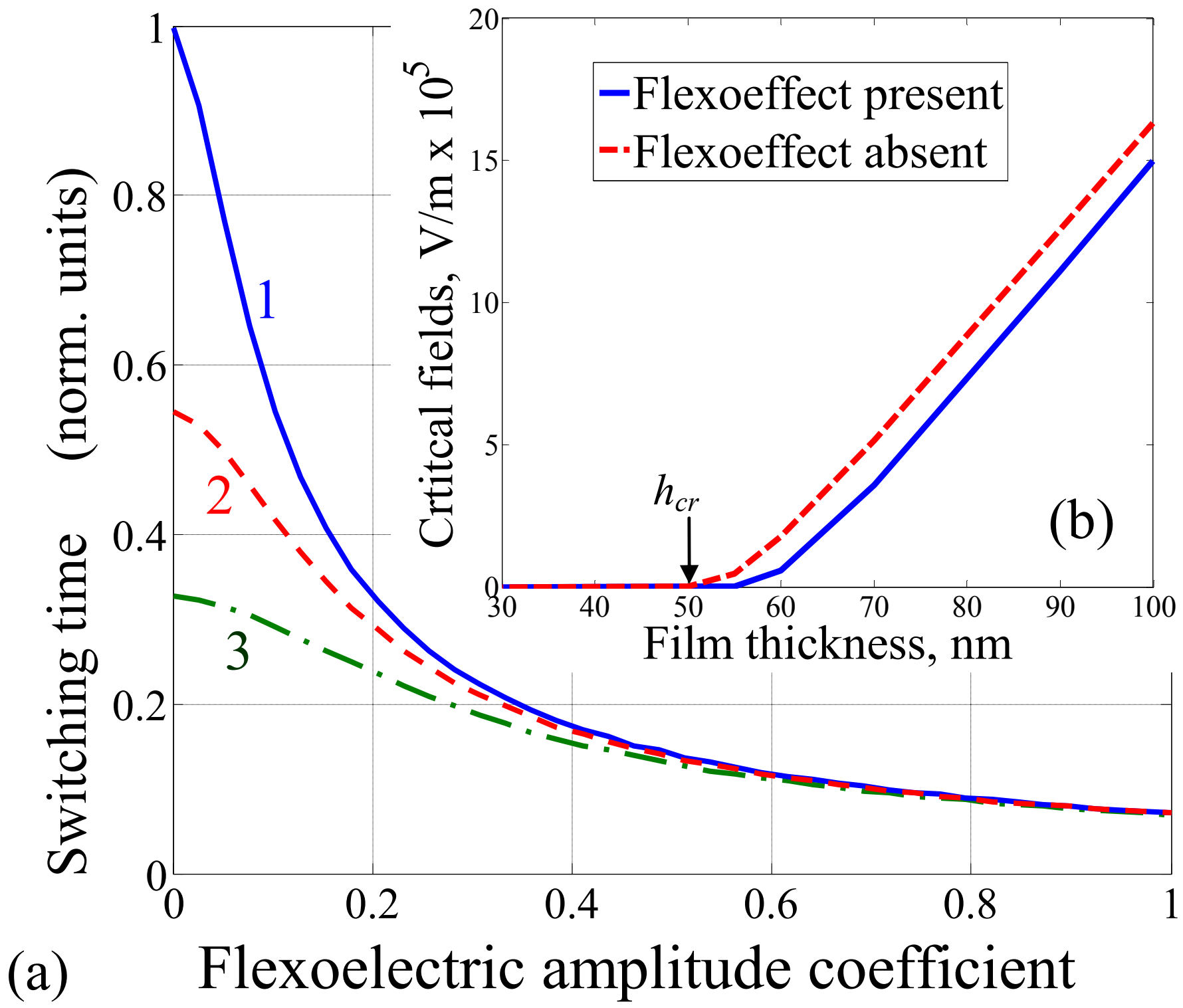

**Figure 5. (a)** Switching time dependence on the related flexoelectric amplitude coefficient FA under the applied electric fields of about 1.631 MV/m (curve 1), 1.633 MV/m (curve 2) and 1.637 MV/m (curve 3). Material parameters are listed in the **Table C1 [44]**. FA = 1 corresponds to the flexocoefficient of $2.46\times10^{-11}$ $m^3/C$, other FA values are the corresponding parts of this value. **Inset (b)** – thickness dependence of the critical field in films with (solid lines) and without (dashed lines) flexoelectric effect.

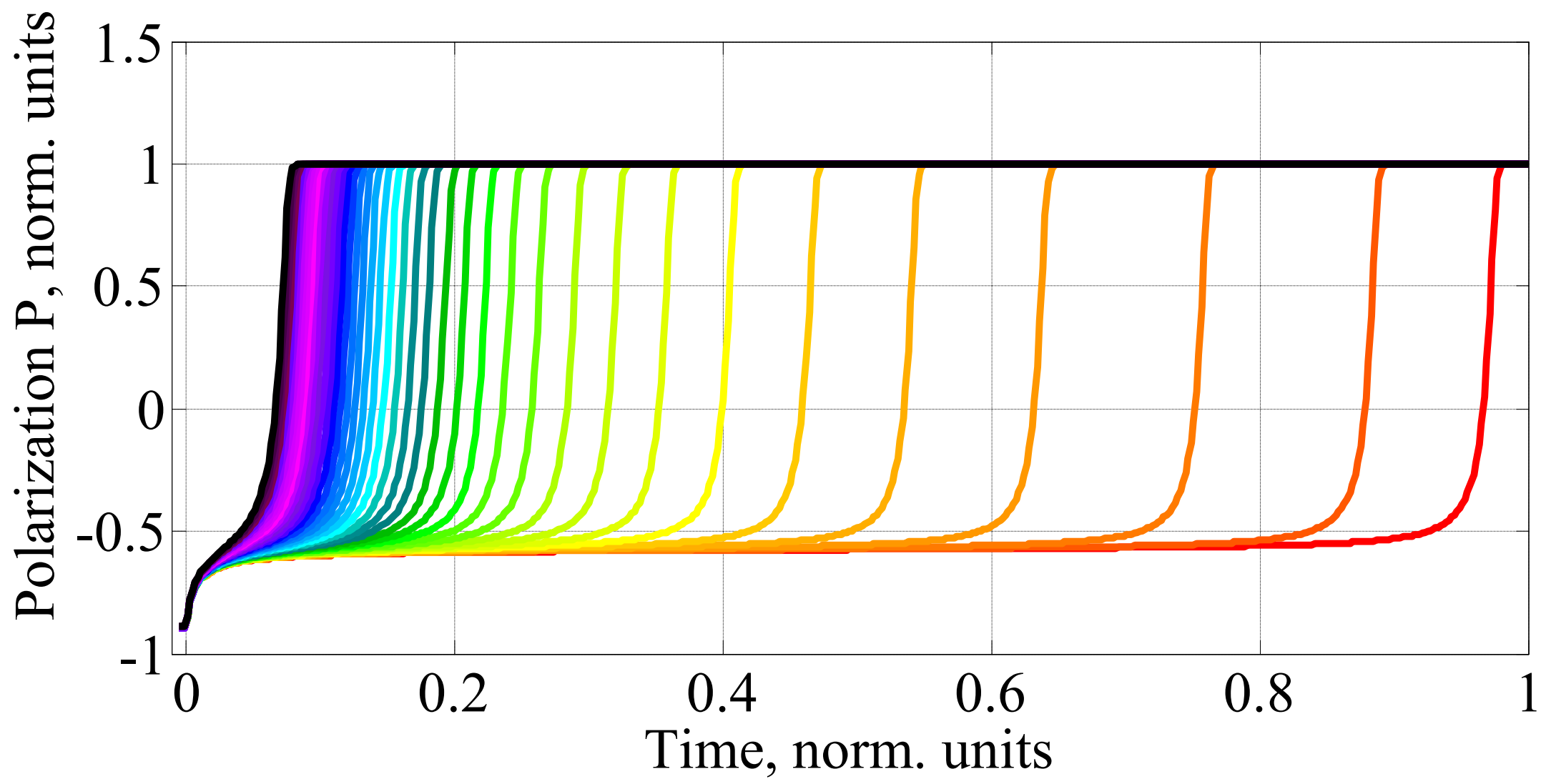

**Figure 6**. Single domain polarization reversal kinetics calculated for the 100 nm film under the applied electric field of 1.631 MV/m. Each curve corresponds to different values of the flexocoupling coefficient defined as FA×$F_{33}$, where $F_{33}$ is the flexoelectric tensor component (see. **Table C1 [44]**) and FA is a relative amplitude coefficient that changes from 0 (red curve at the right) to 1 (black curve at the left), taking 40 equidistant values. The switching time of the system is maximal at FA = 0 (no flexoeffect) and minimal at FA = 1 (flexoeffect coefficient is maximum).

To conclude, the impact of flexoelectric coupling on polarization reversal kinetics and space charge dynamics in thin films of ferroelectric semiconductors was theoretically studied. The Landau-Khalatnikov relaxation equation was used to define a spatial-temporal polarization evolution, together with the Poisson equation for self-consistent determination of electric potential and space charge distribution, as well as the elastic theory equations to determine mechanical stresses, strains and their gradients.

The polarization reversal process consists of two stages for thicker films, while it is reduced to one stage for the thinner films with thickness below the critical value $h_{cr}$. The analysis of obtained results reveals a moderate increase of the flexocoupling influence on the polarization, elastic strain, electric potential and space charge distribution dynamics with the decrease of a ferroelectric film thickness. In contrast, the dependence of polarization switching time on the applied electric field is strongly affected by the flexocoupling strength. Therefore we can conclude that the flexocoupling can affect the dynamic properties of ferroelectric thin films drastically without a significant impact of static distributions. The physical nature of this effect turned out to be related to the strong flexocoupling effect on the distribution dynamics of a polarization gradient and elastic deformation near the film surfaces.

A.N.M. and E.A.E. acknowledge National Academy of Sciences of Ukraine (grant 07-06-15). I.S.V. gratefully acknowledges support from the Deutsche Forschungsgemeinschaft (DFG) through the grant GE 1171/7-1.

48 M. D. Glinchuk, E. A. Eliseev, V. A. Stephanovich, R. Fahri, J. Appl. Phys Vol. 93, № 2, 1150–1159 (2003).

# Supplementary Materials to
# "Flexocoupling impact on the kinetics of polarization reversal"

**Ivan S. Vorotiahin[1, 2], Anna N. Morozovska[1], Eugene A. Eliseev[1, 3], Yuri A. Genenko[21],**

[1] *Institute of Physics, National Academy of Sciences of Ukraine,*

*46, pr. Nauky, 03028 Kyiv, Ukraine*

[2]*Institut für Materialwissenschaft, Technische Universität Darmstadt,*

*Jovanka-Bontschits-Str. 2, 64287 Darmstadt, Germany*

[3] *Institute for Problems of Materials Science, National Academy of Sciences of Ukraine,*

*Krjijanovskogo 3, 03142 Kyiv, Ukraine*

## Appendix A. Elastic fields and Euler-Lagrange equations

For the case of mechanical equilibrium, corresponding equations with $\partial\sigma_{i3}/\partial x_3 = 0$. Boundary conditions corresponds to a mechanically free upper surface of the film, $\sigma_{i3}(0)\big|_S = 0$, and fixed bottom surface z = *h*, where the elastic displacement $u_i$ is zero, i.e. $u_i(h) = 0$. In the case of the film-substrate lattice mismatch presence in-plane strains are fixed at the film-substrate interface, $u_{11}(h) = u_{22}(h) = u_m$.

Elastic field components are quasi-homogeneous [1] for an ultra-thin ferroelectric film (*h*<10 nm) with a constant polarization and without Vegard and flexoelectric effects. In the presence of these effects the approximate expressions for the nonzero components in a very thin film have the following form:

$$\sigma_{11}(x_3) = \sigma_{22}(x_3) = \frac{u_m - Q_{13}P_3^2}{s_{11}+s_{13}} - \frac{W_{11}^d \delta N_d^+}{s_{11}+s_{13}} - \frac{F_{13}}{s_{11}+s_{13}}\frac{\partial P_3}{\partial x_3}, \tag{A.1a}$$

$$u_{11}(x_3) = u_{22}(x_3) = u_m, \qquad u_{33}(x_3) = \frac{2s_{13}u_m}{s_{11}+s_{13}} + Q_{33}^{eff}P_3^2 + W_{33}^{eff}\delta N_d^+ + F_{33}^{eff}\frac{\partial P_3}{\partial x_3}. \tag{A.1b}$$

The apparent coefficients are introduced as $s_{33}^{eff} = s_{33} - \frac{2s_{13}^2}{s_{11}+s_{13}}$, $W_{33}^{eff} = W_{33}^d - \frac{2s_{13}W_{11}^d}{s_{11}+s_{13}}$, $F_{33}^{eff} = F_{33} - \frac{2s_{13}F_{13}}{s_{11}+s_{13}}$ and $Q_{33}^{eff} = Q_{33} - \frac{2s_{13}Q_{13}}{s_{11}+s_{13}}$. Voigt notations are used for the electrostriction $Q_{ij}$, gradient coefficients $g_{ij}$, flexoelectric $F_{ij}$ and elastic compliances $s_{ij}$ tensors, while full matrix notations are used for all other tensors. The tensor components with subscripts 12, 13 and 23 are equal for materials with cubic parent phase.

---

[1] genenko@mm.tu-darmstadt.de

Misfit dislocations, defect concentration gradient and other factors lead to the misfit strain and spontaneous stresses vanishing in thicker films. This becomes clear from the simple energy considerations, because thick strained/stressed films have much higher energy than relaxed ones.

Following Speck and Pompe [2], we modify the solutions (A.1) for a thicker film ($h > h_d$) in the following way:

$$\sigma_{11}=\sigma_{22}=\frac{u_m - Q_{13}P_3^2}{s_{11}+s_{13}}\frac{h_d}{h}-\frac{W_{11}^d\delta N_d^+}{s_{11}+s_{13}}-\frac{F_{13}}{s_{11}+s_{13}}\frac{\partial P_3}{\partial x_3}, \qquad \text{(A.2a)}$$

$$u_{11}=u_{22}=u_m\frac{h_d}{h}, \qquad u_{33}=\left(\frac{2s_{13}u_m}{s_{11}+s_{13}}+Q_{33}^{eff}P_3^2\right)\frac{h_d}{h}+W_{33}^{eff}\delta N_d^+ + F_{33}^{eff}\frac{\partial P_3}{\partial x_3}. \qquad \text{(A.2b)}$$

Here $h_d$ has the sense of some characteristic thickness defined by several factors, such as the critical thickness of misfit dislocation appearance, film thickness-to-width ratio, film-substrate thicknesses ratio, etc.

Actually, it is well-known fact that misfit dislocations emerge in epitaxial films when their thickness is more than the critical thickness of dislocations appearance $h_d$. This thickness decreases with the interface misfit strain $u_m$ increase. In accordance with the Matthews-Blakeslee theory and Speck and Pompe model for perovskites [2], $h_d \approx \frac{b}{u_m}\frac{\sqrt{2}\ln(4h_d/b)}{8\pi(1+\nu)} \sim u_m^{-1}$ in the wide range of $u_m$, which is in good agreement with experiments ($b$ of the order of the lattice constant $a$ is the Burgers vector of dislocation, $\nu$~0.3 is Poisson's ratio). For typical misfit strains $|u_m| \sim 10^{-2}$ the thickness $h_d \sim$ 10–0.5 nm, *i.e.* it is not more than several tens of lattice constants [2]. Hence we can regard that $h_d \le (5-10)$nm. More rigorously, Eqs.(A.2) can be a good approximation in the average sense (as Saint-Venant conditions).

Note, that the piezoelectric contribution is automatically included in the equations (A.1) as linearized electrostriction, since $P_3 \approx P_3^S + \varepsilon_0(\varepsilon_{33}^f - 1)E_3$.

Using Equations (A.1)-(A.2), we exclude the stresses from the equation for polarization $P_3$ and concentration of donors $N_d^+$. Remained equations can be solved numerically. Note, that the substitution of expressions (A.1)-(A.2) in the Gibbs functional leads to the appearance of the flexoelectric coupling with Vegard and piezoelectric terms proportional to the products $W_{ii}^{eff}F_{33}^{eff}$, $\Sigma_{ii}^{eff}F_{33}^{eff}$, etc.

Euler-Lagrange equation for determination of the ferroelectric polarization component has an explicit form:

$$a_{33}^{eff}P_3+b_{33}^{eff}P_3^3+\gamma_{333}P_3^5-g_{33}^{eff}\frac{\partial^2 P_3}{\partial x_3^2}-\frac{2F_{13}W_{11}^d}{s_{11}+s_{13}}\frac{\partial N_d^+}{\partial x_3}=-\frac{\partial\varphi}{\partial x_3} \quad (A.3)$$

The effective coefficients are introduced as $g_{33}^{eff}=g_{33}+\frac{2F_{13}^2}{s_{11}+s_{13}}$, $b_{33}^{eff}=\left(b_{33}+\frac{4Q_{13}^2}{s_{11}+s_{13}}\right)$ and $a_{33}^{eff}=\alpha_{33}^T(T-T_c)-\frac{2u_mQ_{13}}{s_{11}+s_{13}}+\frac{4Q_{13}W_{11}^d}{s_{11}+s_{13}}\delta N_d^+$ for the case $h\le h_d$. For the case $h>h_d$ the coefficients $b_{33}^{eff}=\left(b_{33}+\frac{4Q_{13}^2}{s_{11}+s_{13}}\frac{h_d}{h}\right)$ and $a_{33}^{eff}=\alpha_{33}^T(T-T_c)-\frac{2u_mQ_{13}}{s_{11}+s_{13}}\frac{h_d}{h}+\frac{4Q_{13}W_{11}^d}{s_{11}+s_{13}}\delta N_d^+$.

Boundary conditions for the out-of-plane polarization component $P_3$ are of the third kind [3]:

$$\left.\left(P_3-\lambda_1\frac{\partial P_3}{\partial x_3}-\frac{2F_{13}W_{11}^d}{s_{11}+s_{12}}\frac{\delta N_d^+}{A_{33}^{S1}}-\frac{2F_{13}Q_{13}}{s_{11}+s_{13}}\frac{P_3^2}{A_{33}^{S1}}\right)\right|_{x_3=0}=0, \quad (A.4a)$$

$$\left.\left(P_3+\lambda_2\frac{\partial P_3}{\partial x_3}+\frac{2F_{13}W_{11}^d}{s_{11}+s_{13}}\frac{\delta N_d^+}{A_{33}^{S2}}+\frac{2F_{13}Q_{13}}{s_{11}+s_{13}}\frac{P_3^2}{A_{33}^{S2}}\right)\right|_{x_3=h}=0. \quad (A.4b)$$

The extrapolation length $\lambda_m=\frac{g_{33}^{eff}}{A_{33}^{Sm}}$ is determined by the surface energy and the surface state. Physically realistic range for $\lambda_m$ is 0.5 – 2 nm [4].

## Appendix B

## Coupled equations in dimensionless variables

Poisson equation (S.1) for electric potential $\tilde{\varphi}=e\varphi/k_BT$ in dimensionless variables acquires the form:

$$\frac{\partial^2\tilde{\varphi}}{\partial\tilde{z}^2}=\frac{eP_{s0}L_P}{\varepsilon_0\varepsilon_bk_BT}\frac{\partial\tilde{P}}{\partial\tilde{z}}-\frac{L_P^2}{L_D^2}\left(\tilde{N}(\varphi)-\tilde{n}(\varphi)\right). \quad (B.1)$$

Here we introduced characteristic length scale as a gradient length $L_P=\sqrt{\left(g_{33}+\frac{2F_{13}^2}{s_{11}+s_{12}}\right)\Big/\alpha_{33}^T T_c}$ along with the Debye screening length $L_D=\sqrt{\varepsilon_0\varepsilon_bk_BT/\left(e^2N_s\right)}$, where the equilibrium concentration of ionized donors is $N_s=N_d^0\left(1-f\left((E_d-E_F)/k_BT\right)\right)$. Equation for dimensionless polarization $\tilde{P}=P_3/P_{s0}$ can be found from Eq.(A.3) in the following form

$$\begin{aligned}&\frac{\Gamma}{\alpha_{33}^TT_C}\frac{\partial\tilde{P}}{\partial t}+\left(\frac{T}{T_C}-1+\frac{4Q_{13}W_{11}^dN_s}{\alpha_{33}^TT_C\left(s_{11}+s_{12}\right)}\delta\tilde{N}\right)\tilde{P}+\frac{P_{s0}^2}{\alpha_{33}^TT_C}b_{33}^{eff}\tilde{P}^3+\frac{\gamma_{333}P_{s0}^4}{\alpha_{33}^TT_C}\tilde{P}^5-\\&-\frac{g_{33}+2F_{13}^2/(s_{11}+s_{12})}{\alpha_{33}^TT_CL_P^2}\frac{\partial^2\tilde{P}}{\partial\tilde{z}^2}+\frac{1}{\alpha_{33}^TT_CP_{s0}}\frac{k_BT}{L_Pe}\frac{\partial\tilde{\varphi}}{\partial\tilde{z}}-\frac{2F_{13}W_{11}^dN_s}{\alpha_{33}^TT_CP_{s0}\left(s_{11}+s_{12}\right)L_P}\frac{\partial\tilde{N}}{\partial\tilde{z}}=0\end{aligned} \quad (B.2a)$$

Here we introduced characteristic value of polarization $P_{s0}=\sqrt{\left[\sqrt{\left(b_{33}^{eff}\right)^2-4\gamma_{333}\alpha_{33}^{T}(T-T_c)}-b_{33}^{eff}\right]\Big/2\gamma_{333}}$.

In the dimensionless variables, polarization boundary conditions (4) are transformed to

$$\left.\left(\frac{\tilde{P}}{\tilde{\lambda}}\pm\left(\frac{\partial\tilde{P}}{\partial\tilde{z}}+\frac{2F_{13}W_{11}^{d}N_s}{P_{s0}\left(s_{11}+s_{12}\right)\alpha_{33}^{T}T_cL_P}\delta\tilde{N}+\frac{P_{s0}}{\alpha_{33}^{T}T_cL_P}\frac{2F_{13}Q_{13}}{s_{11}+s_{12}}\tilde{P}^2\right)\right)\right|_{x_3=0,h}=0 \tag{B.2b}$$

Here we introduced a dimensionless extrapolation length as $\tilde{\lambda}=\frac{1}{L_PA_{33}^{S}}\left(g_{33}+\frac{2F_{13}^2}{s_{11}+s_{12}}\right)$.

Dimensionless concentration of donors is

$$\tilde{N}=\frac{N_d^0}{N_s}\tilde{f}\left(\tilde{\varphi}+\tilde{E}_F-\tilde{E}_d+\frac{2W_{11}^{d}}{k_BT\left(s_{11}+s_{12}\right)}\left(W_{11}^{d}N_s\delta\tilde{N}+\frac{F_{13}P_{s0}}{L_P}\frac{\partial\tilde{P}}{\partial\tilde{z}}+Q_{13}P_{s0}^2\tilde{P}^2\right)\right). \tag{B.3}$$

Also we will need a dimensionless concentration of electrons, which can be expressed as:

$$\tilde{n}=\frac{N_C}{N_s}F_{1/2}\left(\tilde{\varphi}+\tilde{E}_F-\tilde{E}_C\right) \tag{B.4}$$

Boundary conditions for electric potential are

$$\left.\tilde{\varphi}\right|_{\tilde{z}=0}=\tilde{V},\quad \left.\tilde{\varphi}\right|_{\tilde{z}=\tilde{h}}=0. \tag{B.5}$$

Dimensionless out-of-plane displacement at the surface is

$$\left.\frac{u_3}{L_P}\right|_{\tilde{z}=\tilde{h}}=\left(\tilde{F}_{33}-\frac{2s_{13}\tilde{F}_{13}}{s_{11}+s_{12}}\right)\left(\left.\tilde{P}\right|_{\tilde{z}=\tilde{h}}-\left.\tilde{P}\right|_{\tilde{z}=0}\right)+\left(\tilde{Q}_{33}-\frac{2s_{13}\tilde{Q}_{13}}{s_{11}+s_{12}}\right)\int_0^{\tilde{h}}\tilde{P}^2d\tilde{z}\quad (\text{at } h\le h_d) \tag{B.6a}$$

$$\left.\frac{u_3}{L_P}\right|_{\tilde{z}=\tilde{h}}=\left(\tilde{F}_{33}-\frac{2s_{13}\tilde{F}_{13}}{s_{11}+s_{12}}\right)\left(\left.\tilde{P}\right|_{\tilde{z}=\tilde{h}}-\left.\tilde{P}\right|_{\tilde{z}=0}\right)+\frac{h_d}{h}\left(\tilde{Q}_{33}-\frac{2s_{13}\tilde{Q}_{13}}{s_{11}+s_{12}}\right)\int_0^{\tilde{h}}\tilde{P}^2d\tilde{z}\quad (\text{at } h>h_d) \tag{B.6b}$$

Dimensionless variables and parameters involved in Eqs.(B.1)-(B.6) are listed in the **Table B1.**

In the dimensionless LGD equation Landau-Khalatnikov coefficient $\Gamma$ becomes $\tilde{\Gamma}$. Standing in front of the time derivative this coefficient is measured in the time units. The factor $\frac{1}{\alpha_{33}^{T}T_C}$ determines the unit on time scale, which in this case is equal to ~$10^{-8}$ s. $\tilde{\Gamma}$ is assigned a value of $10^{-3}$ s to observe the whole switching process within the given time interval (~10 ns), which corresponds to the characteristic order of $\Gamma$ ~$10^{6}$ s.

**Table B1. Dimensionless variables and parameters**

| **Quantity** | **Definition/designation** |
|---|---|
| Gradient length – characteristic value of length | $L_P=\sqrt{\left(g_{33}+\frac{2F_{13}^2}{s_{11}+s_{12}}\right)\Big/\alpha_{33}^{T}T_c}$ |
| Debye screening length | $L_D = \sqrt{\varepsilon_0 \varepsilon_b k_B T / (e^2 N_s)}$ |
| coordinate and thickness | $\tilde{z} = x_3 / L_P$, $\tilde{h} = h / L_P$ |
| donor concentration | $\tilde{N} = \frac{N_d^0}{N_s} \tilde{f}\left( \begin{array}{l} \tilde{\varphi} + \tilde{E}_F - \tilde{E}_d + \\ + \frac{2W_{11}^d}{k_B T(s_{11} + s_{12})}\left( \begin{array}{l} W_{11}^d N_s \delta\tilde{N} + \tilde{F}_{13} \frac{\partial \tilde{P}}{\partial \tilde{z}} + \\ + \tilde{Q}_{13} \tilde{P}^2 \end{array} \right) \end{array} \right)$ |
| electron density | $\tilde{n} = \frac{N_C}{N_s} F_{1/2}\left( \tilde{\varphi} + \tilde{E}_F - \tilde{E}_C \right)$ |
| Equilibrium concentration of ionized donors at zero potential and stress | $N_s = N_d^0 \left( 1 - f\left( (E_d - E_F) / k_B T \right) \right)$ |
| electric potential | $\tilde{\varphi} = e\varphi / k_B T$ or $\varphi = \tilde{\varphi} \frac{k_B T}{e}$ |
| electric field | $\tilde{E}_3 = L_P e E_3 / k_B T$ or $E_3 = \tilde{E} \frac{k_B T}{L_P e}$ |
| Applied voltage | $\tilde{V} = eU / k_B T$ |
| chemical potential of electrons | $\tilde{\zeta}_d = \zeta_d / k_B T$ |
| donor level and conduction band | $\tilde{E}_d = E_d / k_B T$  $\tilde{E}_C = E_C / k_B T$ |
| Fermi level | $\tilde{E}_F = E_F / k_B T$ |
| polarization | $\tilde{P} = P_3 / P_{s0}$ or $P_3 = P_{s0} \tilde{P}$ |
| Characteristic value of polarization | $P_{s0} = \sqrt{\left[ \sqrt{\left(b_{33}^{eff}\right)^2 - 4\gamma_{333} \alpha_{33}^T (T - T_c)} - b_{33}^{eff} \right] / 2\gamma_{333}}$ |
| Vegard coefficient | $\tilde{W} = W_{11}^d N_s$ |
| Dimensionless concentration | $\tilde{n}_b = \frac{n_b}{N_s}$ |
| Electrostriction coefficient | $\tilde{Q}_{13} = Q_{13} P_{s0}^2$, $\tilde{Q}_{33} = Q_{33} P_{s0}^2$ |
| flexoelectric coefficient | $\tilde{F}_{13} = F_{13} \frac{P_{s0}}{L_P}$, $\tilde{F}_{33} = F_{33} \frac{P_{s0}}{L_P}$ |
| Extrapolation length | $\tilde{\lambda} = \frac{1}{L_P A_{33}^S} \left( g_{33} + \frac{2F_{13}^2}{s_{11} + s_{12}} \right)$ |
| Landau-Khalatnikov coefficient | $\tilde{\Gamma} = \frac{\Gamma}{\alpha_{33}^T T_C}$ |

## Appendix C

**Figures C1-C2** show the static distributions of polarization, electrostatic potential, donor and electron concentrations for the thick (100 nm) and thin (10 nm) ferroelectric films with and without flexoelectric effect.

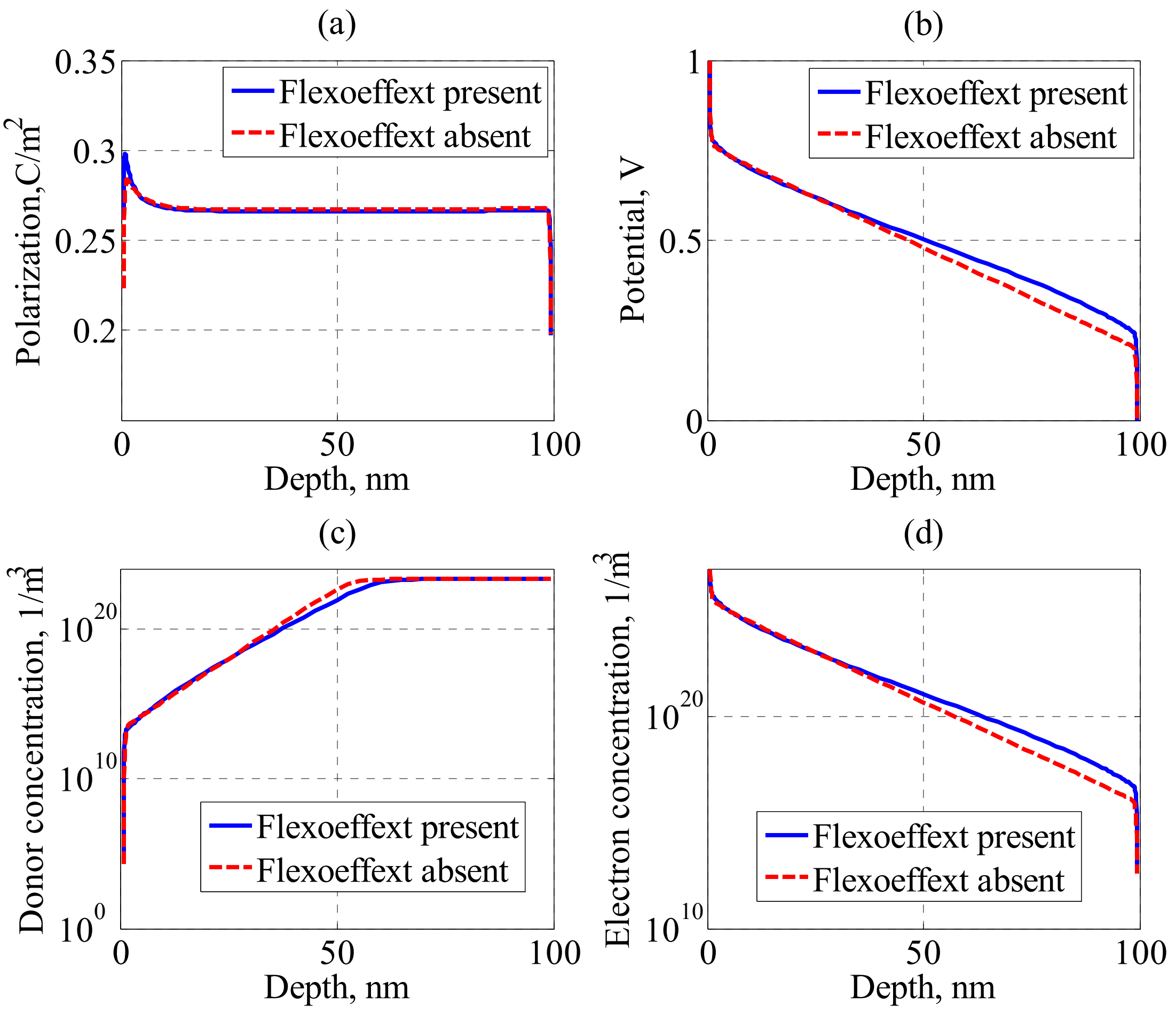

**Figure C1.** Spatial distributions of the polarization **(a)**, electrostatic potential **(b)**, donor **(c)** and electron **(d)** concentrations for the film of thickness 100 nm with (blue solid curve) and without (red dashed curve) flexoelectric effect. The voltage applied to the film is 1 V. Material parameters are listed in **Table C1**.

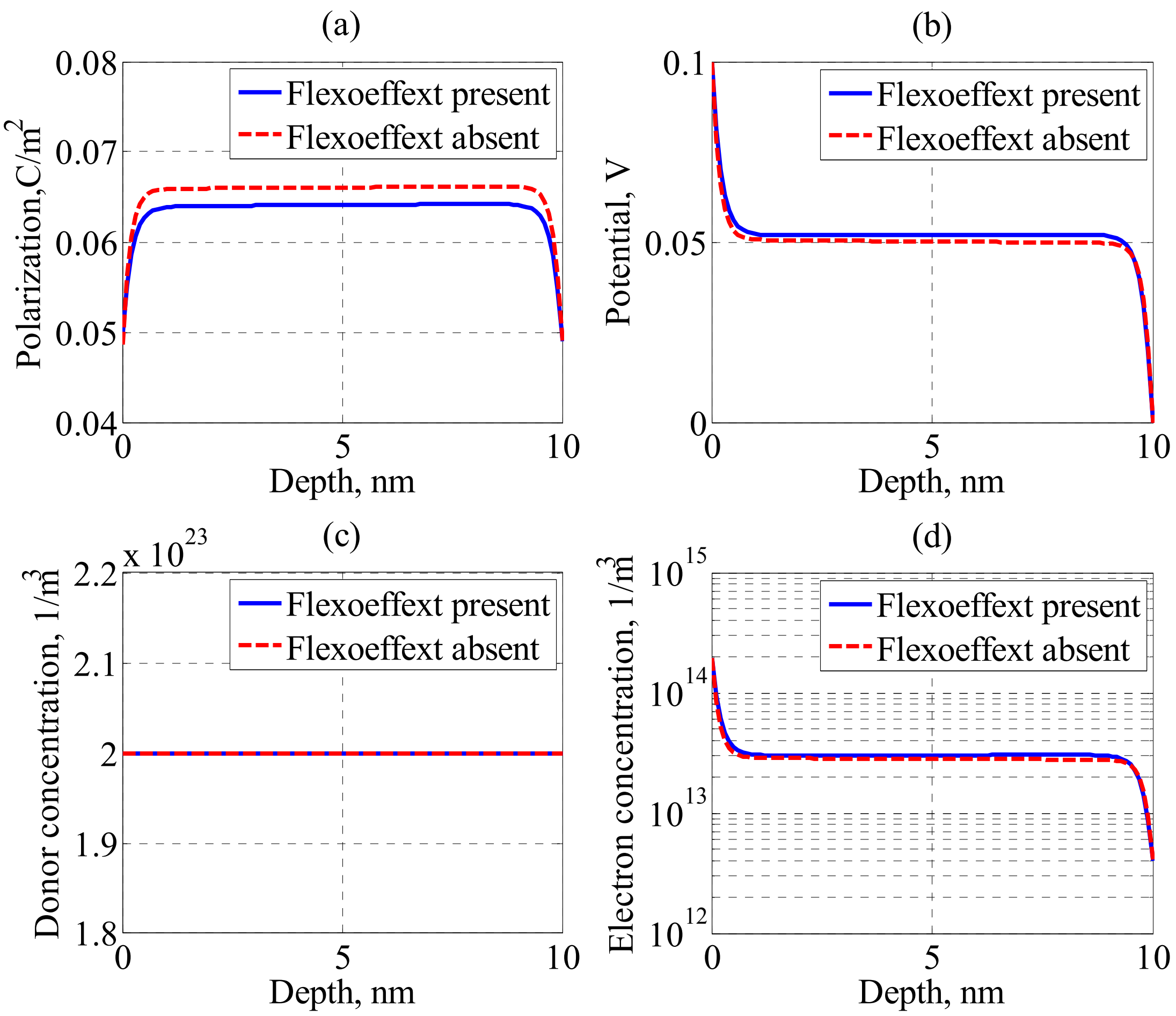

**Figure C2.** Spatial distributions of the polarization **(a)**, electrostatic potential **(b)**, donor **(c)** and electron **(d)** concentrations for the film of thickness 10 nm with (blue solid curve) and without (red dashed curve) flexoelectric effect. The voltage applied to the film is 0.1 V. Material parameters are listed in **Table C1**.

The distributions similar to those shown in figures C1 and C2 can also be obtained for the films of various thicknesses between 100 and 10 nm. Note that one can observe the changes in the distribution behaviour when changing the film thickness. With the decrease of the film thickness the flexocoupling influence on the static distributions increases, the critical fields $E_{cr}$ necessary for the polarization reversal fall dramatically, amplitudes of the spontaneous and field-induced polarizations decrease moderately. In the film of thickness 47 nm the spontaneous polarization amplitude becomes negligibly small in comparison with the value of 0.19 C/m² for the 100 nm film, . Below 47 nm the reversal process changes, as the spontaneous polarization distribution allows the film to switch in one step, because the spontaneous polarization distribution is close to the homogeneous one in the equilibrium state.

Thus, it is clearly seen, that the relative difference between the distributions calculated for the films with and without flexoeffect increases with the film thickness decrease. The Vegard effect contribution appears to be very small, so that the distributions for films with and without Vegard effect barely differ. Asymmetry of the physical variable distributions under the applied electric field is caused by the field itself given its relatively high amplitude, and is also related to the different distributions of electrons and donors. Furthermore, mechanical boundary conditions also contribute into the charge distributions in the equation system (see Appendix A, B), because mechanical properties are specified through the modified coefficients of the flexocoupling and Vegard effect (see **Appendix B**). As the electron concentration distribution follows the electrostatic potential, it can be assumed that applied field disturbs the equilibrium, of the charge carriers and donors, and thus causes inhomogeneities in the physical variable distributions.

In the absence of applied electric field, the system has a stable energetically favorable state with the spontaneous polarization. Polarization distribution throughout the film in this case is linear and monotonous in the film bulk with abrupt drops toward zero near the film surfaces (see **Figure C3a**). Electrostatic potential creates two areas in the film bulk with positive and negative values respectively and drops to zero at the surfaces, to obey the boundary conditions. Potential in the bulk obtains maximal absolute values near the surfaces, just before the film boundaries, and linearly goes to the opposite maximum at the other surface, crossing zero (see **Figure C3 b**). Flexocoupling impacts these distributions by breaking their symmetry.

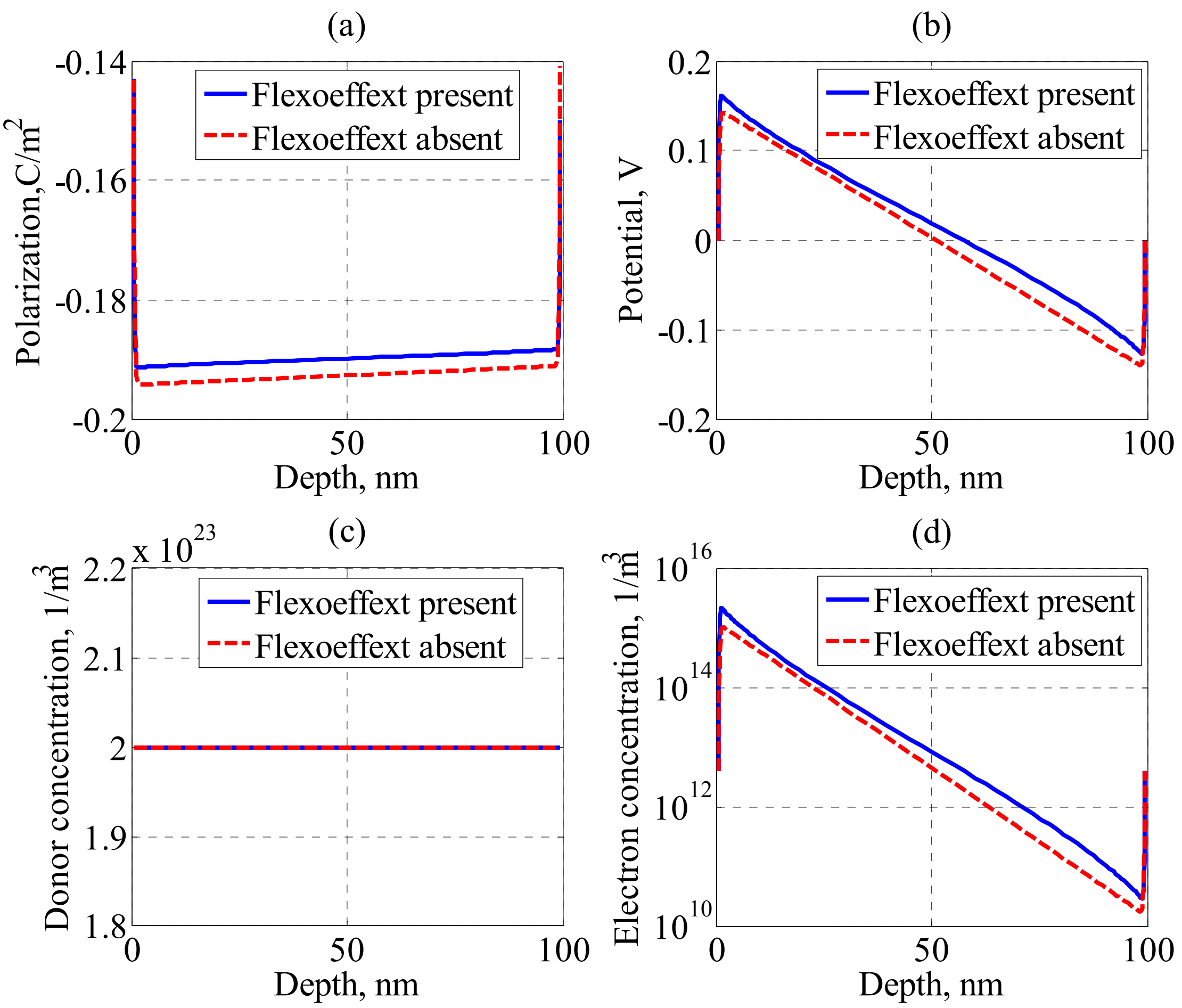

**Figure C3.** Spontaneous polarization **(a)** and electric potential **(b)** distributions in the 100 nm film without applied electric field. Material parameters are listed in **Table C1**. Blue solid line represent distribution calculated with account for the flexoelectric and Vegard effects, red dashed line represents the distribution without Vegard effect, green dash-dotted line denotes the absence of flexocoupling, and the black dotted line describes distributions without any of those effects.

**Figures C4** and **C5** show the kinetics of polarisation switching in the form of time dependence of polarization together with displacement currents as first time derivatives of polarization. It is seen that a small increase of electric field, as well as a small increase of flexoelectric strength, can significantly decrease the switching time under certain conditions (i.e. at the applied fields close to the $E_{cr}$).

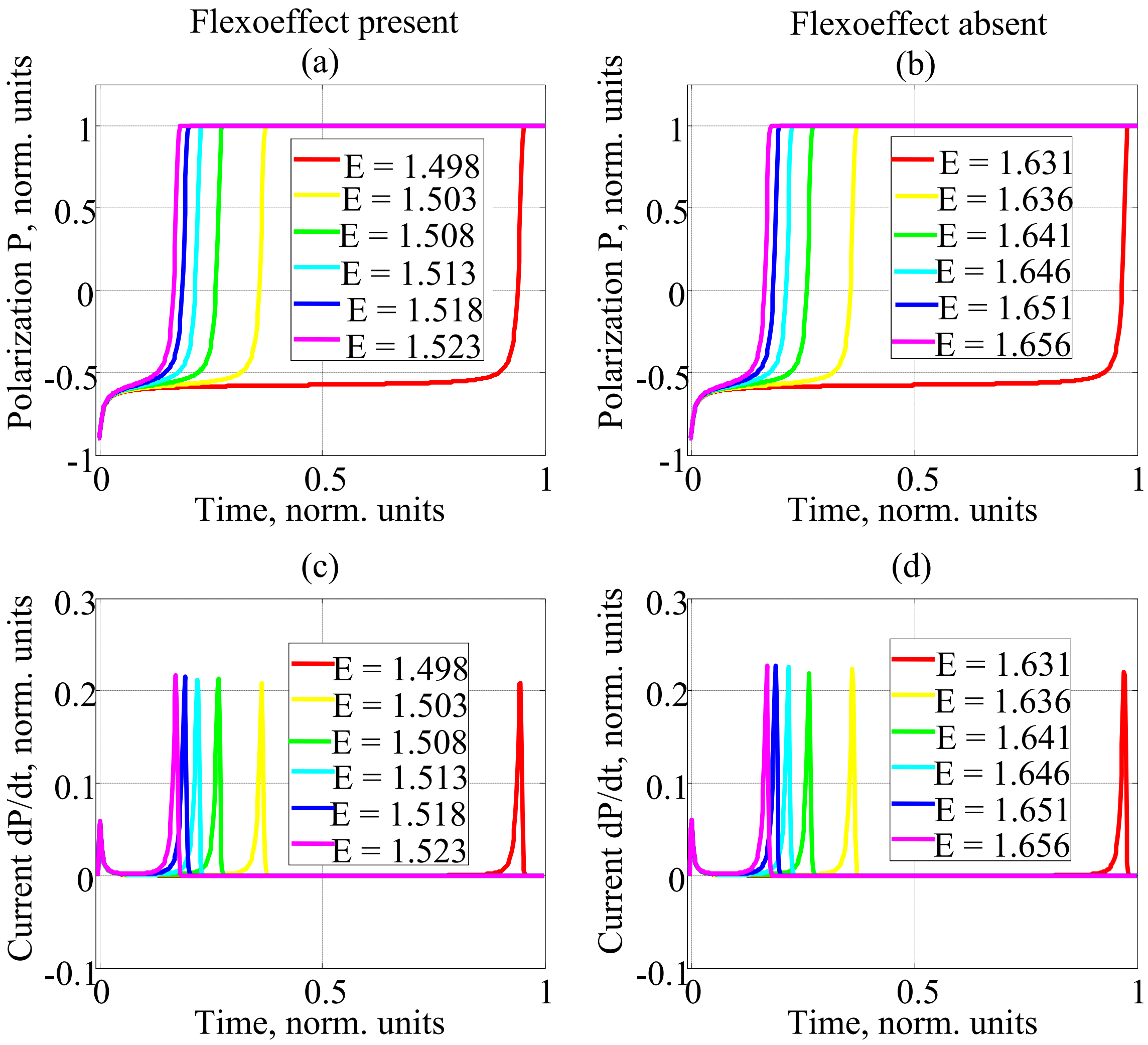

**Figure C4.** Single domain polarization reversal kinetics (a, b) and correspondent displacement currents (c,d) calculated for the 100 nm films with **(a, c)** and without **(b, d)** flexoelectric effect. Applied electric field is measured in MV/m. Parameters used in the modeling are listed in **Table C1**.

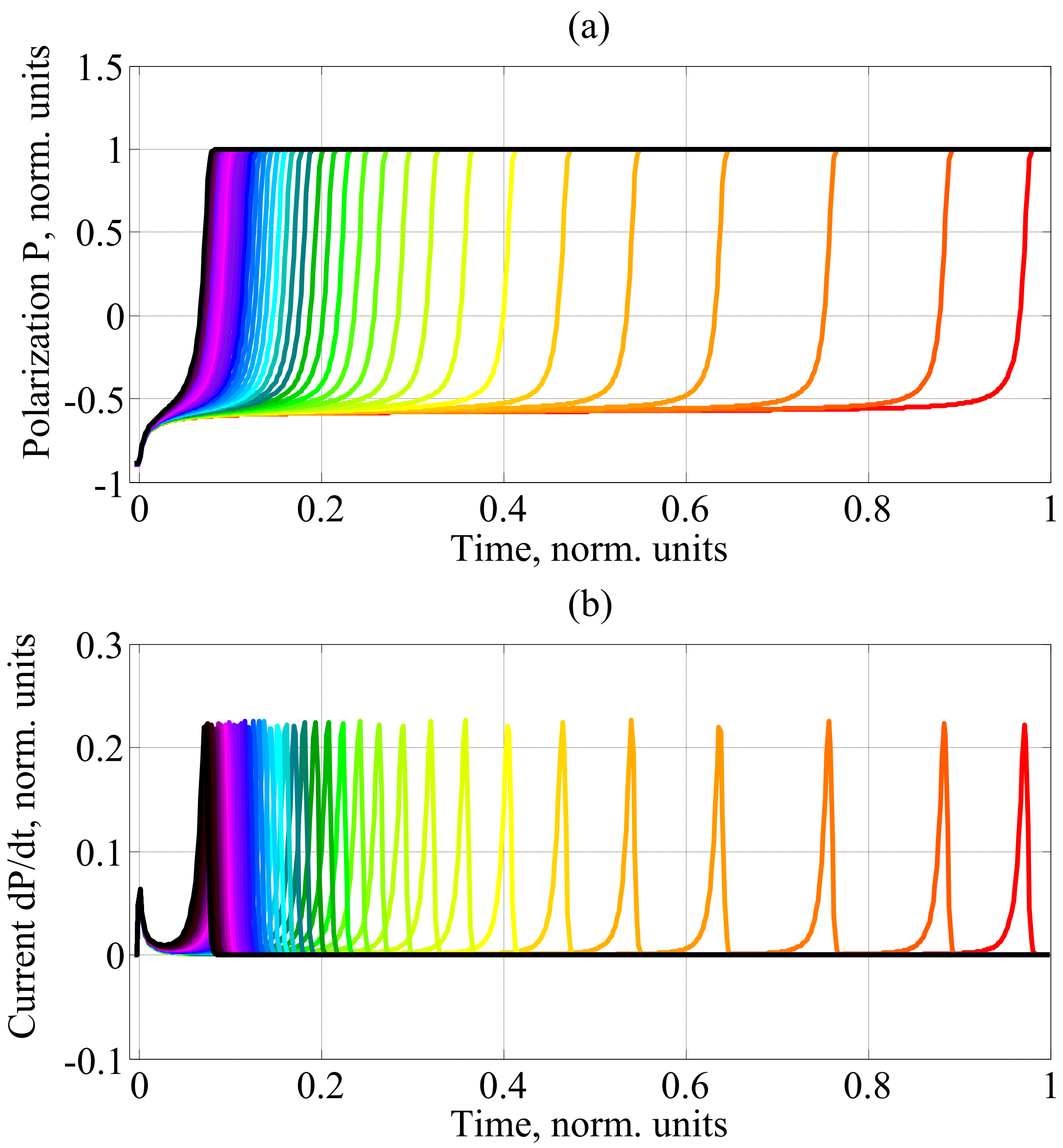

**Figure C5**. Single domain polarization reversal kinetics **(a)** and correspondent displacement current **(b)** calculated for the 100 nm film under the applied electric field of 1.631 MV/m. Each curve corresponds to different values of the flexocoupling coefficient defined as FA×$F_{33}$, where $F_{33}$ is the flexoelectric tensor component (see. **Table C1**) and FA is a relative amplitude coefficient that changes from 0 (red curve at the right) to 1 (black curve at the left), taking 40 equidistant values. The switching time of the system is maximal at FA = 0 (no flexoeffect) and minimal at FA = 1 (flexoeffect coefficient is maximum).

**Table C1. Known material properties and experimental parameters**

| | |
|---|---|
| Material | $BaTiO_3$ |
| Structure | Tetragonal |
| Temperature | 300°K |
| Curie Temperature | 400°K |
| Film thickness *h* | 100 nm (75, 60, 45, 30, 10) |
| Background dielectric permittivity $\varepsilon_b$ | 7 |
| Vegard expansion tensor *w* | 10 Å³ |
| Effective electron mass $m_n$ | $0.3m_e$ |
| Electron mass $m_e$ | $9.11 \cdot 10^{-31}$ kg<br>$8.19 \cdot 10^{-14}$ J<br>$0.511 \cdot 10^{6}$ eV |
| $E_c$ - $E_d$ | 0.1 eV |
| $E_c$ | 0.85 eV |
| $E_d$ | 0.75 eV |
| $E_F$ | 0 eV |
| Predicted spontaneous polarization $P_S$ | 0.27 C/m² |
| Maximum of donor concentration $N_d^0$ | $10^{23}$ 1/m³ |
| ***LGD and elastic tensors:*** | |
| $a_{33}$ | $-2.94 \cdot 10^{7}$ mJ/C² (m/F) |
| $\alpha_{33}^{T}$ – temperature independent coefficient | 3.34 $10^{5}$ |
| $b_{33}$ | $-6.71 \cdot 10^{8}$ $m^5J/C$ |
| $b_{12}$ | $3.23 \cdot 10^{8}$ $m^5J/C$ |
| $\gamma_{333}$ | $8.004 \times 10^{9}$ ($C^{-6}m^9J$) |
| $\gamma_{112}$ | $4.47 \times 10^{9}$ ($C^{-6}m^9J$) |
| $\gamma_{123}$ | $4.91 \times 10^{9}$ ($C^{-6}m^9J$) |
| $g_{33}$ | $5.1 \cdot 10^{-10}$ $m^3J/C$ |
| $g_{12}$ | $-0.2 \cdot 10^{-10}$ $m^3J/C$ |
| $g_{44}$ | $0.2 \cdot 10^{-10}$ $m^3J/C$ |
| $s_{11}$ | $8.3 \cdot 10^{-12}$ 1/Pa |
| $s_{13}$ | $-2.7 \cdot 10^{-12}$ 1/Pa |
| $s_{44}$ | $9.24 \cdot 10^{-12}$ 1/Pa |
| $c_{11}$ | $0.12 \cdot 10^{12}$ Pa |
| $c_{13}$ | $-0.37 \cdot 10^{12}$ Pa |
| $c_{44}$ | $0.108 \cdot 10^{12}$ Pa |
| $Q_{33}$ | 0.11 $m^4/C^2$ |
| $Q_{13}$ | $m^4/C^2$<br>-0.043 |
| $Q_{44}$ | $m^4/C^2$<br>0.059 |
| $F_{33}$ | $2.46 \cdot 10^{-11}$ m³/C |
| $F_{13}$ | $0.48 \cdot 10^{-11}$ m³/C |
| $F_{44}$ | $0.05 \cdot 10^{-11}$ m³/C |
| ***Recalculated values*** ($F_{ij} = f_{ik} \cdot s_{jk}$; $Q_{ij} = q_{ik} \cdot s_{jk}$) | |
| $f_{33}$ | 2.96 J/C |
| $f_{44}$ | 0.054 J/C |
| $q_{33}$ | $13.2\cdot10^{9}$ J·m/C² |
| $q_{44}$ | $6.372\cdot10^{9}$ J·m/C² |
| | |
| ***Additional parameters*** | |
| Donor charge $Z_d$ | 2 |
| Electrochemical potential of electrons $\zeta_e$ | Equal to Fermi level |
| Electrochemical potential of donors $\zeta_d$ | Equal to Fermi level |
| Landau-Khalatnikov coefficient $\Gamma$ | $1.336\times10^{6}$ sec. |
| Evaluation of the Khalatnikov term $\Gamma/(\alpha_t T_c)$ | $10^{-3}$ |
| Inhomogeneous movable species concentration $\delta N_d^+$ | Coordinate-dependent |
| Depolarization distance $h_d$ | 10 nm |
| Extrapolation length $\lambda$ | 0.5 nm |
| Misfit $u_m$ | 0 |

## Appendix D

## Fermi integral and its Approximation

To solve Poisson's equation for electrostatic potential the numerical calculation of the space charge concentration is required. The electron concentration is strongly dependent on the potential itself and positions of the energy levels in bulk, i.e. bottom of the conduction band in relation to the Fermi level. For calculation the Fermi integral $F_{1/2}(\xi)=\frac{2}{\sqrt{\pi}}\int_0^{\infty}\frac{\sqrt{\zeta}d\zeta}{1+\exp(\zeta-\xi)}$ should be used as the reflection of Fermi-Dirac statistics for the electron distribution.

The above integral is a form of a polylogarithm, so that no explicit analytical solution can be shown. Instead we need to calculate this integral numerically, using an approximation. A useful expression for the approximate Fermi integral was presented by X. Aymerich-Humet et al. [5,6], having the following form:

$$\tilde{F}_j(x)=\frac{1}{\Gamma(j+1)}\left(\frac{(j+1)\cdot 2^{j+1}}{\left[b+x+\left(|x-b|^{c}+a^{c}\right)^{1/c}\right]^{j+1}}+\frac{e^{-x}}{\Gamma(j+1)}\right)^{-1}, \tag{D.1}$$

where $j$ is the order of integral, $\Gamma(n)$ is the Gamma-Euler-function, a factorial function with the following properties: $\Gamma(n)=(n-1)!$, $\Gamma(1/2)=\sqrt{\pi}$, $\Gamma(p+1)=p\Gamma(p)$. *a, b, c* denote the short forms of the polynomial expressions $a = \left[ 1 + \frac{15}{4}(j+1) + \frac{1}{40}(j+1)^2 \right]^{\frac{1}{2}}$, $b = 1.8 + 0.61 \cdot j$, $c = 2 + (2 + \sqrt{2}) \cdot 2^j$.